\documentclass[manuscript,nonacm]{acmart}
\usepackage{graphicx}
\usepackage{amsmath}
\usepackage{booktabs}
\usepackage{subcaption}
\usepackage{pdflscape}
\usepackage{adjustbox}
\usepackage{geometry}
\usepackage{listings} % For code listings

\title{Working Document - Formalising Software Requirements with Large Language Models}
\author{Arshad Beg}
\authornotemark[1]
\authornote{All authors contributed equally to this research.}
\email{arshad.beg@mu.ie}
\orcid{0009-0004-6939-0411}
\author{Diarmuid O'Donoghue}
\authornotemark[1]
\email{diarmuid.odonoghue@mu.ie}
\orcid{0000-0002-3680-4217}
\author{Rosemary Monahan}
\authornotemark[1]
\email{rosemary.monahan@mu.ie}
\orcid{0000-0003-3886-4675}
\affiliation{%
  \institution{Maynooth University}
  \city{Maynooth}
  \country{Ireland}
}
\begin{document}

\begin{abstract}

This draft is a working document, having a summary of nighty-four (94) papers with additional sections on Traceability of Software Requirements (Section \ref{sec:traceability}), Formal Methods (Section \ref{sec:FMnTools}), Unifying Theories of Programming (UTP) and Theory of Institutions (Section \ref{sec:UTPnToInst}). Please refer to abstract of \cite{Beg2025, Beg2025a}. \textbf{Key difference} of this draft from our recently anticipated ones with similar titles, i.e. AACS 2025 \cite{Beg2025} and SAIV 2025 \cite{Beg2025a} is: 

\cite{Beg2025} is a two page submission to ADAPT Annual Conference, Ireland. Submitted on 18th of March, 2025, it went through the light-weight blind review and accepted for poster presentation. Conference was held on 15th of May, 2025.

\cite{Beg2025a} is a nine page paper with additional nine pages of references and summary tables, submitted to Symposium on AI Verification (SAIV 2025) on 24th of April, 2025. It went through rigorous review process. The uploaded version on arXiv.org \cite{Beg2025a} is the improved one of the submission, after addressing the specific suggestions to improve the paper. 

%The suggestions are listed below:
%
%Taxonomy. 
%group the 25 papers (prompt-only, fine-tuned, verifier-in-loop, etc.) \\
%add a one-page comparative table (accuracy, human effort, domain). \\
%Performance ceiling. Which tasks show the highest reliability today (assertion generation vs. full contracts)? Summarise trends. \\
%Trim the Dafny lemma. Keep a short illustrative snippet; move full proof to appendix.\\

\end{abstract}


%% Keywords

%\received{24 February 2025}
%\received[revised]{12 December 2025}
%\received[accepted]{5 June 2026}

\maketitle

\section{Introduction}

Relying on natural language in software requirements  often leads to ambiguity. In addition, requirements which are not expressed in a formal mathematical notation cannot be  guaranteed through formal verification techniques as required to meet standards e.g. \cite{10.1145/2070336.2070341,6136916,10.5555/1151816.1151817} in safety critical software. Expressing requirements in formal notation requires training in the domain of requirements engineering, as well as knowledge of formal notation and associated proof methods, often increasing the software development cycle time by a factor of 30\% \cite{Huisman2024}. In this project, we aim to ease the burden of writing specification, helping to bridge the gap between the need for formal verification techniques and their lack of use in the software industry (due to the fast pace of the industry environment). 

Formalising software requirements ensures clarity, correctness and verifiability, and requires formal specification languages, logic and verification techniques, such as theorem proving and model-checking, to guarantee the correctness of the software system under construction. Like all other fields, the development of Large Language Models (LLMs) has opened a world of opportunities where we can exploit their power to generate formal requirements and accompanying specifications. 

In this paper, we present the results of our structured literature review which examines how large language models are currently used to assist in writing formal specifications. Main research questions for conducting systematic literature review on the topic are as follows: \\
\textbf{RQ1:} What methodologies leverage Large Language Models (LLMs) to transform natural language software requirements into formal notations? \\
% while addressing the associated challenges and limitations identified by researchers? - [addressed in section \ref{literature_formalisation_llms}]\\
\textbf{RQ2:} What are the emerging trends and future research directions in using LLMs for software requirements formalisation? 
%- [addressed in section \ref{sec:future_directions} with discussion in succeeding sub-section \ref{subsec:CoTnPromptEngg}]

The structure of the paper is organized as follows. Section \ref{sec:methodologyforsurvey} presents the methodology used to select the relevant literature. Section \ref{literature_formalisation_llms} provides a brief overview of works focused on the formalisation of software requirements using large language models (LLMs) addressing \textbf{RQ1}. Literature concerning the traceability of software requirements is discussed in Section \ref{sec:traceability}. Section \ref{sec:FMnTools} lists down well-known formal notations developed over the last three decades of research by the formal method community and its related tools. Section \ref{sec:UTPnToInst} explores studies involving formal proofs within the frameworks of Unifying Theories of Programming (UTP) and the Theory of Institutions. Section \ref{sec:future_directions} presents future directions based on our findings and the discussion on chain of thought and prompt engineering in sub-section \ref{subsec:CoTnPromptEngg}, addressing \textbf{RQ2}. Concluding remarks are provided in Section \ref{sec:conclusions}.

\section{Methodology for Literature Review}
\label{sec:methodologyforsurvey}
To conduct a structured and thorough review of literature on Natural Language Processing (NLP), Large Language Models (LLMs), and their use in software requirements, the following approach is followed. Several academic databases, including IEEE Xplore, ACM Digital Library, Scopus, Springer Link, and Google Scholar, are searched using specific keywords. The core search terms include “NLP,” “LLMs,” and “Software Requirements,” with broader terms such as “specification,” “logic,” “verification,” “model checking,” and “theorem proving” used to expand the scope. The number of results differs notably across databases. For example, IEEE Xplore returns 17 peer-reviewed articles, Scopus lists 20, Springer Link filters to 595, ACM Digital Library provides 1,368 results, and Google Scholar shows 14,800 references since 2021. These discrepancies highlight the importance of applying precise selection methods to extract the most relevant studies.

%\dodnote{Thinking as a review: they will ask who says Elicit is a good tool for this? Why is it better than the previous approach? Do we need to write defensively here? Can one of our results be an evaluation of Elicit? (ratio of relevant papers)}

 %\rmnote{what was the criteria used for inclusion, exclusion etc.}

To streamline the process of locating strong contributions, the AI-powered tool Elicit \cite{elicit_tool} is used. Elicit supports the literature review by offering summarised content and DOIs for suggested papers. While it helps reduce the manual workload during the initial phase, every suggested paper in this review is manually reviewed to ensure its relevance. This ensures that the final list excludes any unrelated or off-topic material. After the initial filtering, a manual review is performed to confirm the relevance and quality of each paper. Abstracts are first assessed to judge suitability. If the abstract lacks clarity or depth, a further examination of the full text is conducted.
%Elicit has shown a good level of accuracy in delivering relevant research results, making the search process smooth and generally reliable. In our experience, every time a provided DOI was accessed, it led to a legitimate, peer-reviewed research paper, which reflects well on the tool’s consistency. 
% While it's important to remain cautious with any AI tool, Elicit has so far provided a dependable experience, especially for those looking to explore academic content with a balance of convenience and credibility.
%Although Elicit assists with the early stages of reference gathering, the final decisions remain in the hands of the researchers. For each section of this survey, Elicit is employed only for identifying references, maintaining a clear balance between automation and expert judgement.

%After the selection, a detailed manual evaluation is carried out to confirm that each chosen paper provides substantial input to the field. 
Abstracts are closely read, and when necessary, the full text is reviewed using the following exclusion and inclusion criteria:

%This ensures that only papers offering significant value to the discussion on NLP, LLMs, and software requirements are included. From the larger collection of retrieved literature, selected works %are shortlisted based on their contribution, citation count, and alignment with the aims of this review. These are presented in Section \ref{literature_formalisation_llms}. Manual checks guarantee %that all included articles offer strong theoretical or empirical foundations, and that no redundant or weak contributions are carried over.
%\abnote{Inclusion and Exclusion Criteria}
%Based on above methodology, we sketch exclusion and inclusion criteria explicitly as follows:

\textbf{Inclusion Criteria:} Studies are included if they offer meaningful theoretical or empirical insights related to NLP, LLMs, and their application in software requirements. This includes topics like specification, formal logic, verification, and formal methods. 

\textbf{Exclusion Criteria:} Papers are excluded if they show in-sufficient relevance to the intersection of NLP/LLMs and software requirements, or if their abstracts or full texts lack sufficient detail. Non-peer-reviewed materials, duplicates, and items suggested by Elicit but deemed irrelevant after manual review are also removed.
%The methodology balanced automation with human judgment, ensuring that the final set of papers captured critical advancements in the field. The survey not only relies on published peer-reviewed research but also follows an iterative approach where insights from the reviewed literature guide further keyword adjustments and additional searches. This iterative process enhances the comprehensiveness of the survey, ensuring a well-rounded perspective on the state-of-the-art in NLP, LLMs, and software requirements.

\section{Formalising Requirements through LLMs}
\label{literature_formalisation_llms}
The paper \cite{10500073} proposes using LLMs, like GPT-3.5, to verify code by analysing requirements and explaining whether they are met. The work \cite{cosler2023nl2specinteractivelytranslatingunstructured} details about nl2spec, a framework that leverages LLMs to generate formal specifications from natural language, addressing the challenge of ambiguity in system requirements. Users can iteratively refine translations, making formalization easier. The work \cite{cosler2023nl2specinteractivelytranslatingunstructured} provides an open-source implementation with a web-based interface.

The work \cite{quan2024verificationrefinementnaturallanguage} provides verification and refinement of natural language explanations by making LLMs and theorem provers work together. A neuro-symbolic framework i.e. Explanation-Refiner is represented. LLMs and theorem provers are integrated together to formalise explanatory sentences. The theorem prover then provides the guarantee of validated sentence explanations. Theorem prover also provides feedback for further improvements in NLI (Natural Language Inference) model. Error correction mechanisms can also be deployed by using the tool Explanation-Refiner. Consequently, it automatically enhances the quality of explanations of variable complexity.  \cite{arora2023advancingrequirementsengineeringgenerative} outlines key research directions for the stages of software requirement engineering, conducts a SWOT analysis, and share findings from an initial evaluation. 

In \cite{10207159}, symbolic NLP and ChatGPT performance is compared while generating correct JML output against given natural language pre-conditions. In \cite{10.1145/2976767.2976769}, domain models are generated against natural language requirements for industry-based case study. Domain model extractor is designed and applied to four industrial requirements documents. The accuracy and overall performance is reported for the designed model extractor. An automatic synthesis of software specifications is provided through LLMs in \cite{mandal2023largelanguagemodelsbased}. We know that software configurations give an insight of how the software will behave. While software are frequently discussed and documented in a variety of external sources, including software manuals, code comments, and online discussion forums, making it hard for system administrators to get the most optimal configurations due to lack of clarity and organisation in provided documentation. Work \cite{mandal2023largelanguagemodelsbased} proposed SpecSyn framework that uses an advanced language model to automatically generate software specifications from natural language text. It treats specification generation as a sequence-to-sequence learning task and outperforms previous tools by 21\% in accuracy, extracting specifications from both single and multiple sentences. 

AssertLLM tool is presented in \cite{10691792}. The tool generates assertions to do hardware verification from design specifications, exploiting three customised LLMs. It is done in three phases, first understanding specifications, mapping signal definitions and generating assertions. The results show that AssertLLM produced 89\% correct assertions with accurate syntax and function. The work \cite{10.1002/spe.430} reports on formal verification of NASA's Node Control Software natural language specifications. The software is deployed at International Space Station. Errors found in the natural language requirements are reported by the authors with a commentary on lessons learnt.

SpecLLM \cite{li2024specllmexploringgenerationreview} explores the space of generating and reviewing VLSI design specifications with LLMs. The cumbersome task of chip architects can be improved by exploiting the power of LLMs for synthesising natural language specifications involved in chip designing. So, the utility of LLMs is explored with the two stages i.e. (1) \textbf{generation} of architecture specifications from scratch and from register transfer logic (RTL) code; and (2) \textbf{reviewing} these generated specifications. 

The paper \cite{nowakowski2013requirements} introduced a model-based language (Requirements Specification Language - RSL) that enhances software requirements specification by incorporating constrained natural language phrases, including verbs, adjectives, and prepositions, grouped by nouns. It also presented an advanced tooling framework that captures application logic specifications, enabling automated transformations into code, validated through a controlled experiment. The framework functionality is integrated as development platform named ReDSeeDS. The work was a part of EU project, maintained at www.redseeds.eu \cite{nowakowski2013requirements}. A similar work is reported in \cite{Ghosh2016} which describes the details of \verb"ARSENAL" framework and methodology designed to perform automatic requirements specification extraction from natural language. The generated specification can be verified automatically through the framework. 

An interesting work is presented in \cite{6823180}. Business process models specifes the requirements of process-aware information systems. It includes generation of natural language from business process models. Domain experts and system analysts find it difficult to validate BPMs directly. The automation of generating a text which describes these models is done in \cite{6823180}. For the translation process, BPMs are translated to RPST - a tree like structure of the process model which is then used to generate sentences first. These generated sentences are refined by adding linguistic complexity later-on. The generated natural language is found complete and more understandable. In primitive work of 1996 \cite{491451}, the software requirements were expressed in a limited set of natural language. Authors \cite{491451} referred it as controlled natural language. For Controlled Language (CL) , authors used the Alvey Natural Language Toolkit (ANLT). This intermediate notation is then translated to logical expressions in order to detect and remove ambiguities in requirement specifications. In \cite{reinpold2024exploringllmsverifyingtechnical}, the potential and power of LLMs is exploited for smart grid requirement specifications improvement. Here, the performance of GPT-4o and Claude 3.5 Sonnet is analysed through f1-scores, achieving in range of 79\% - 94\%. 

The paper \cite{10684640} reports the translation between NL and Linear Temporal Logic (LTL) formulas through the use of LLMs. The challange of low accuracy and high cost during model training and tuning of general purpose LLMs is considered in  \cite{10684640}. Dynamic prompt generation and human interaction with LLMs are amalgamated to deal with the mentioned challenges. Unstructured natural language requirements are converted to NL-LTL pairs. The approach achieved up to 94.4\% accuracy on publicly available datasets with 36 and 255,000 NL-LTL pairs. Dealing with a different domain domain, it improved from 27\% to 78\% through interactive prompt evolution.

In \cite{tihanyi2024newerasoftwaresecurity}, authors present ESBMC-AI, a framework that combined Large Language Models (LLMs) with Formal Verification to automatically detect and fix software vulnerabilities. The approach used Bounded Model Checking (BMC) to identify errors and generate counterexamples, which were then fed into LLM to repair the code, followed by re-verification with BMC. Evaluated on 50,000 C programs from the FormAI dataset, ESBMC-AI effectively fixed issues like buffer overflows and pointer dereference failures with high accuracy, making it a valuable tool for software development and CI/CD integration. In \cite{hahn2022formalspecificationsnaturallanguage}, authors reported performance of LLMs to translate natural language into formal rules by training them on datasets for regular expressions (regex), first-order logic (FOL), and linear-time temporal logic (LTL) . The results show that the models adapt well to new terms and symbols, and outperformed existing methods in regular expression translation.

In \cite{mukherjee2024automatedverificationllmsynthesizedc}, SynVer is presented. SynVer is a framework to synthesise and verify C programs. Built on the Verified Software Toolchain, the tool usage was applied on benchmarks containing basic coding tasks, Separation Logic assertions, and API specifications. In \cite{10.1145/3643991.3644922}, five different benchmark datasets are used to analyse the performance of GPT-3.5 and GPT-4, making sure the low-level software requirements meet all high-level requirements. The results reported in \cite{10.1145/3643991.3644922} showed that GPT-3.5, using zero-shot prompting with explanations, correctly detected full coverage in four out of five datasets and achieved 99.7\% recall in spotting missing coverage from a removed low-level requirement.

SAT-LLM, a unique framework to remove conflicting requirements is represented in \cite{10.1145/3691620.3695302}. It integrated Satisfiability Modulo Theories (SMT) solvers with LLMs. The performance of LLMs struggles while removing complex requirements conflicts. With the capability of formal reasoning, integrating SMT solvers with LLMs makes it a viable approach for the task. Experiments are performed with SAT-LLM and it performed well as compared to standalone performance of LLM i.e. ChatGPT identified 33\% of conflicts with a Precision of 0.85, Recall of 0.31, and an F1 score of 0.46, struggling with hidden or complex conflicts. SAT-LLM performed better, identifying 80\% of conflicts with a Precision of 1.00, Recall of 0.83, and an F1 score of 0.91. \cite{ernst:LIPIcs.SNAPL.2017.4} used input of error messages, variable names, procedure documentation and user questions. The suitablility of NLP techniques is discussed by going through the literature for each step of software development life cycle. The authors of \cite{ernst:LIPIcs.SNAPL.2017.4} indicated NLP a good candidate for software development. For instance, they discussed available literature for generating assertions by synthesising sentences in testing phase.  

The paper \cite{Req2SpecPaper} introduced Req2Spec, an NLP-based tool that analyses natural language requirements to create formal specifications for \verb"HANFOR", a large-scale requirements and test generation tool. Tested on 222 automotive software requirements at BOSCH, it correctly formalized 71\% of them. A primitive work in the domain is about RML \cite{Greenspan1986}. RML bundled with features of writing requirements, which are based on conceptual model, having attribute of organisation and abstraction, maintaining precision, consistency and clarity. Well-defined logic and semantics of RML are presented in \cite{Greenspan1986}. 
The work \cite{fan2025evaluatingabilitylargelanguage} evaluated GPT-4o's ability to generate specifications for C programs that can be verified using VeriFast, a static verifier based on separation logic. Their experiments, which use different user inputs and prompting techniques, show that while GPT-4o's specifications maintain functional behaviour, they often fail verification and include redundancies when verifiable. The primitive work \cite{Nelken1996} described automatic translation from natural language sentences to temporal logic, in order to deploy formal verification of the requirements.

Paper \cite{wu2024lemurintegratinglargelanguage} introduced a methodology Lemur that integrated large language models with automated reasoners for program verification, formally defining transition rules, proving their soundness, and demonstrating practical improvements on synthetic and competition benchmarks. Performance of Lemur is compared with Code2Inv, ESBMC, and UAutomizer while using SV-COMP benchmarks. Code2Inv is a comprehensive work of 2020 \cite{Si2020}, contributing towards learning-based end-to-end program verification. Code2Inv utilised reinforcement learning to train an invariant synthesizer to propose loop invariants. \cite{Necula2024} is a systematic review of the domain i.e. translating between natural language software requirements to formal ones, based on the databases of Scopus, ACM, IEEE Xplore and Clarivate. \cite{10628478} proposed a pipeline integrating Large Language Models (LLMs) to automatically refine and decompose safety requirements for autonomous driving, addressing frequent updates in the automotive domain. The work evaluated LLMs' ability to support Hazard Analysis and Risk Assessment (HARA) through iterative design science and expert assessments, ultimately implementing the prototype in an industrial setting. The responsible software development team evaluated its efficiency. 

\cite{7092662} presented a framework to ensure consistency between different representations of specifications, maintaining semantic alignment between oral and formal descriptions while ensuring implementation potential through synthesis. It included time extraction, input-output partitioning, and semantic reasoning beyond syntactic parsing. The framework of \cite{7092662} performed well on various test examples. The work \cite{10.5220/0006817205010512} reports available NLP tools for Topological Functioning Modelling (TFM). Among six selected LLM pipelines, best performance tools found were the Stanford CoreNLP toolkit, FreeLing, and NLTK toolkit. In \cite{10.1145/3643763}, power of LLMs is utilised for generating Dafny tasks. 178 problems are selected from MBPP benchmark. Three types of prompts are used: (1) Context less prompts, (2) Signature prompt comprised of method signature and test cases, and (3) Context of Thought (CoT) prompt based on decomposition of problems into multiple steps and inclusion of retrieval augmentation generated problems and solutions. GPT-4 outperformed PaLM-2 on the evaluated tasks and achieved the best results using the retrieval-augmented CoT prompt. The technique described in \cite{10.1145/3643763} provided 153 verified Dafny solutions to MBPP problems, including 103 synthesized by GPT-4 and 50 written manually. 

The work \cite{Yang2023} introduced LeanDojo, an open-source toolkit that enables programmatic interaction with Lean theorem prover, providing annotated proof data to support premise selection in theorem proving. Using LeanDojo’s extracted data, \cite{Yang2023} developed ReProver, a retrieval-augmented LLM-based prover that effectively selects premises, significantly improving theorem proving efficiency while requiring only one GPU week of training. Furthermore, a new benchmark with 98,734 theorems and proofs from Lean’s math library constructed. It is designed to test generalization to novel premises, and demonstrate ReProver’s superiority over non-retrieval baselines and GPT-4. Thor \cite{Jiang2022} is a framework developed to integrate language models with theorem provers. The task of finding relevant premises while proving conjecture is the crucial task in implementing automatic theorem provers. Thor \cite{Jiang2022} is presented to handle this very task. The class methods for selecting relevant premises are named Hammers. To test the performance of the framework, the PISA dataset is used. It improved the accuracy from 39\% to 57\%, while 8.2\% of the proofs could not be solved neither from the language models nor from the theorem provers. Thor also outperformed the previous works while using MiniF2F dataset. The paper \cite{Granberry2025} explores the space of integrating Co-pilot (github) with formal methods. It introduces key formal languages i.e. Dafny, Ada/SPARK, Frama-C, and KeY alongwith the interactive theorem provers (Coq, Isabelle / HOL and Lean). The integration of Copilot and formal methods is proposed through development of IDE containing language servers. The examples of such existing IDEs are VSCode and Eclipse where multiple programming languages support is available in one IDE through Language Server Protocol (LSP). 

A very comprehensive study on the state of the art of formal specification and verification on autonomous robotic systems is \cite{10.1145/3342355}. The authors did literature review on cyber-physical systems, omitting pure software-based systems. The survey covered human-controlled and autonomous systems. These were either remotely operated or self-governing. Formal properties like safety, security and reliability were considered. Mechanical and physical considerations are excluded from the scope of the survey \cite{10.1145/3342355}.

\cite{Casadio2025} is a comprehensive work on NLP verification, where existing verification approaches are exhaustively analysed in the paper. A structured \textbf{NLP Verification Pipeline} has been established compromising six critical components: data selection, generation of perturbations, choice of embedding functions, definition of subspaces, robust training, and verification through existing algorithms. To validate this pipeline, ANTONIO tool has been implemented. It enabled modular experimentation with various pipeline components. The key contribution of \cite{Casadio2025} is the identification of gaps in existing approaches and the proposal of novel solutions to improve the robustness and reliability of NLP verification pipelines. NLP verification results have been reported through additional criteria. Standard verifiability metrics has been extended, comparing geometric with sematic subspaces. Semantic perturbations are employed while conducting experiments. In \cite{Casadio2025}, the importance of reporting volumes, generalisability, and embedding error of verified subspaces is emphasised. The reason behind is the great impact of these factors on reliability and interpretability of verification results. The paper \cite{Casadio2025} indicates the future expansion by evaluating model robustness against adversarial perturbations and dataset variations. 

Granberry et al. \cite{Granberry2025a} explored how combining large language models (LLMs) with symbolic analysis can help generate specifications for C programs. They enhanced LLM prompts using outputs from PathCrawler and EVA to produce ACSL annotations. Their findings showed that PathCrawler generated context-aware annotations, while EVA contributed to reducing runtime errors. 

The purpose of Dafny is to automate proofs by outsourcing them to an SMT solver. The SMT solver needs assertions while automating the process. \cite{mugnier2024laurelgeneratingdafnyassertions} presented a framework named Laurel to generate Dafny assertions using LLMs. Mugnier et al. \cite{mugnier2024laurelgeneratingdafnyassertions} designed two domain-specific prompting techniques. First one locates the position in code where assertion is missing. This is done through analysis of the verifier's error message. At the particular location with missing assertion, a placeholder is inserted. Second technique involves provision of example assertions from codebase. Laurel was able to generate over 50\% of the required helper assertions, making it a viable approach to deploy, while automating program verification process. 

The work \cite{Ma2024} represents a novel framework named SpecGen to generate specifications through LLMs. Two phases are applied. First phase is about having prompts in conversational style. Second phase is deployed where correct specifications are not generated. Here, four mutation operators are applied to ensure the correctness of the generated specifications. Two benchmarks i.e. SV-COMP and SpecGen are used. Verifiable specifications are generated successfully for 279 out of 384 programs, making \cite{Ma2024} a viable approach. The work \cite{BoraCaglayan2024} deals with the challenges involved in NL2SQL transformation, being widely deployed in Business Intelligence (BI) applications. Bora et al. \cite{BoraCaglayan2024} developed a new benchmark focused on typical NL questions in industrial BI scenarios. Authors added question categories in the developed benchmark. Furthermore, two new semantic similarity evaluation metrics are represented in \cite{BoraCaglayan2024}, increasing NL2SQL transformation capabilities. 

\section{Traceability of Software Requirements}
\label{sec:traceability}

A dynamic requirements traceability model is proposed in \cite{1618501}, enabling improved software quality through verification and validation of functional requirements. The model ensured software scalability as well, dealing both small and large-sized projects. A novel model for traceability and verification in the early development phase is presented in \cite{Salem2010}. Adaptability to requirement changes is improvised in the model and its impact is assessed. \cite{Genvigir2010} provided a comprehensive review of software requirements traceability, covering key elements, challenges, and techniques. The study classified traceability approaches and highlighted prospects for future research.

The empirical analysis of requirements completeness is performed in \cite{7723818}. By enforcing completeness in requirements ensured reduced defect rates. Here, traceability metrics is produced and regression analysis is performed to quantify the software quality. \cite{Hayes2007} introduced the RETRO tool for automating requirements traceability matrix (RTM) generation. The study showed that RETRO significantly improved accuracy and efficiency compared to manual tracing methods. \cite{doi:10.1142/S0218194020500278} proposed a hybrid approach combining VSM and BTM-GA to enhance traceability link generation. The method outperformed traditional IR techniques, improving recall and precision, particularly in agile development contexts.

Trustrace \cite{6341764} is a trust-based traceability recovery approach. It leverages mined data from software repositories. In comparison to standard IR techniques, Trustrace improved precision and recall in traceability link retrieval. \cite{10.1109/ICSE.2017.9} applied deep learning for traceability, incorporating semantic understanding and domain knowledge. This is done using BI-GRU. It outperformed traditional approaches of VSM and LSI in terms of accuracy and effectiveness. A topology-based model-driven approach for backward requirements traceability is presented in \cite{DBLP:conf/enase/AsninaGOADS11}, formalising specifications and establishing trace links between real-world functional units and software artifacts.

In order to manage software evolution process, \cite{1232285} proposed an event-based traceability mechanism. The work reported performance improvement in change management by linking artifacts through an event service. It maintained consistency being deployed in distributed development environment. Tracebok \cite{7765529} is a body of knowledge on software requirements traceability. The framework categorised traceability approaches and provided guidance for implementing traceability in software projects. \cite{10.1007/978-981-16-8129-5_7} reviewed visualisation tools and techniques for software requirements traceability. The challenges emphasised were scalability and visual cluttering, providing insights of improved traceability visualisation.

In \cite{4550895}, Z notation is used to represent the SRS and design artifacts in order to establish the traceability of functional requirements. The SRS document originally used UML diagrams for requirement analysis and software design. Z notation established trace paths based on defined rules. A prototype framework based on XML is developed to trace the requirements in software design. The designed framework is named as RVVF (Requirement Verification and Validation Framework).  \cite{Cerbah2001} established the concept that terminology extraction can improve traceability from formal models to textual requirements. Cerbah et al. \cite{Cerbah2001} presented a fully implemented system that analysed text corpora to generate hierarchically organised terminological resources and formal class hierarchies. These resources and formal class hierarchies are then communicated to the Troeps knowledge server via an XML stream. The system also enabled user validation of terminology elements and supports bidirectional linkage between the model and source documents. Survey paper \cite{TORKAR2012} examined requirements traceability, including definitions, challenges, and tools. \cite{Pinheiro1996} represented a tool named TOOR - traceability of object-oriented requirements. The tool is based on the principles of hyper-programming and hyper-requirements.   
 
Goknil et al. \cite{Goknil2011} addressed requirements and their relationships from a traceability perspective. It introduced a metamodel for requirements that included formally defined relation types. The relations are formalised using first-order logic to enable inference and consistency checking. A supporting tool demonstrated the approach on a real-world requirements document, helping to uncover hidden dependencies and detect contradictions.

\section{Formal Methods and Testing}
\label{sec:FMnTools}

A detailed examination of the interaction between formal specification techniques and software testing is presented in the 2009 ACM survey \cite{HBB09}. The paper argues that formal descriptions of systems can significantly support the testing process. It introduces a classification of formal specification languages into several distinct categories. These include model-based methods such as Z and VDM, languages grounded in finite-state representations like FSMs and Statecharts, algebraic approaches such as OBJ, process algebraic notations like CSP and CCS, and hybrid methods that integrate both continuous and discrete system behaviours—although the latter are treated as outside the scope of the survey.

In practical applications, formal methods can contribute to testing by either enabling the automated generation of test cases or offering mechanisms to define precise test oracles. Executable specifications, in particular, may be subjected to model-checking techniques to assess conformance with desired properties. The theoretical underpinnings that link formal specifications with testing processes are also explored, with \cite{Gau95} offering foundational contributions—especially in identifying test selection assumptions that underpin the effectiveness of test generation. Each category of formalism is associated with its own techniques for producing relevant test artefacts.

\subsection{Model-Based}

In this category, testing often involves partitioning the input domain using assumptions about uniform system behaviour within each partition. Logical expressions—frequently cast in disjunctive normal form—are used to define these partitions and guide automation. Further approaches include domain analysis techniques that focus on identifying critical boundaries for functions and operators. Additionally, refinement-based testing and mutation techniques are discussed as part of this landscape. Although highly suitable for defining test oracles, model-based methods are often limited in their ability to automatically generate tests without the assistance of theorem-proving tools.

\subsection{Finite State-Based}

Testing approaches that rely on finite-state representations frequently define correctness in terms of language-based conformance between the implementation and its specification. A common strategy involves using a fault model to constrain the number of system states considered. Test generation techniques initially focus on deterministic finite-state machines, before expanding to address partial and non-deterministic forms. While these methods offer structured ways to derive test suites, the primary challenge lies in managing the combinatorial growth in possible state sequences.

\subsection{Process Algebras}

Systems described using process algebra are typically interpreted through labelled transition systems (LTS), which can be infinite. To address this, state space reduction techniques are often employed. In many respects, the methods and challenges of testing in this domain resemble those found in finite-state approaches, particularly with regard to scalability and complexity management.

\subsection{Algebraic}

Algebraic specifications are especially well-suited to object-oriented software. Test cases in this setting may be derived either from the syntactic structure of operations or from the logical axioms that define their intended behaviours. This method provides a strong formal basis for validation, although transforming abstract axioms into executable test procedures requires further elaboration and interpretation.

The survey places particular emphasis on the value of automated reasoning tools in supporting test activities. Model checkers are highlighted as being capable of producing counterexamples when temporal properties are not satisfied, which can subsequently be re-purposed as test cases. Similarly, properties defined in temporal logic can guide the construction of structured test sequences. 

\subsection{Formal Tools}

Isabelle/HOL is a powerful theorem prover based on higher-order logic. It features robust automation, a large repository of verified theorems, and tools for interactive proof construction. The system allows formal reasoning about complex mathematical properties and software systems, offering both depth and flexibility in its approach.

Frama-C is a modular platform designed for the formal analysis of C code. It supports ACSL (ANSI-C Specification Language) for specifying expected behaviour and includes plug-ins for static analysis, verification, and integration with theorem provers. This makes it highly applicable to domains that demand rigorous validation of software correctness.

There has been significant progress in the use of probabilistic techniques for formal verification \cite{KNP02, HeraultMP04, FGT11}. Probabilistic model-checking focuses on evaluating how likely a system is to meet certain criteria, rather than delivering absolute verdicts. A comprehensive overview of developments in this area is provided in \cite{AP18}, particularly in the field of statistical model-checking, which offers scalable solutions for analysing stochastic systems in practical contexts.

Promela is a modelling language aimed at the representation of concurrent systems. It is paired with the SPIN model checker, which is widely used in both academic and industrial settings. A supporting tool, Modex, can automatically extract Promela models from C code. SPIN evaluates properties defined in linear-time temporal logic (LTL), and when verification fails, the counterexamples it provides can be used to create test cases. Its command-line interface makes it suitable for automation and integration into larger tool-chains.

TLA+ offers a mathematical framework for specifying systems, particularly those involving concurrency. It emphasises logical precision using simple mathematical constructs. PlusCal provides a more familiar, algorithm-like syntax that compiles directly into TLA+, enabling a smoother transition for developers accustomed to pseudocode-style representations.

\section{Formal Proofs in UTP and Theory of Institution}
\label{sec:UTPnToInst}

\cite{Woodcock2004} provided a tutorial introduction to Hoare and He's Unifying Theories of Programming (UTP) and the concept of designs. It explained how alphabetised relational calculus could describe various programming constructs, illustrating their application to imperative programming theories like Hoare logic and the refinement calculus. \cite{10.1145/147508.147524} introduced the concept of institutions as a formal framework to model logical systems. It presented several foundational results, such as gluing signatures, preserving theory structuring, and extending institutions to include constraints for abstract data types, contributing significantly to the theory of specifications and programming languages.

\cite{Woodcock2002} discussed Circus, a concurrent language that integrated imperative programming, CSP, and Z through the unifying theories of programming. It provided a formalisation of Circus in the UTP framework, highlighting its use for refining concurrent systems. \cite{DBLP:journals/isse/HammerCHJR22} explored integrating runtime verification into an automated UAS Traffic Management system. It demonstrated how runtime verification could ensure system safety by applying formal requirements to various subsystems, validated through real-world flight simulations.

\cite{DBLP:conf/birthday/ButterfieldT23} presented formal verification applied to the RTEMS real-time operating system, using Promela models and the SPIN model-checker to verify multi-core processor qualification for spaceflight. It discussed linking UTP semantics to enhance the test generation process, with a focus on future research directions.

\cite{FOSTER2020102510} introduced Isabelle/UTP, an implementation of Hoare and He's UTP for unifying formal semantics. It enabled mechanising computational theories across paradigms and provided proof tools for Hoare logic, refinement calculus, and other computational paradigms, supporting the development of automated verification tools. \cite{6058982} discussed the use of UML and the UML Testing Profile (UTP) in model-based testing for resource-constrained real-time embedded systems. It addressed the generation of test artefacts from UTP standards and presented a detailed algorithm for creating test cases for such systems.

\cite{Gadia2016} presented a Java model of the priority inheritance protocol in the RTEMS real-time operating system, verified using Java Pathfinder. It detected and fixed known bugs in the RTEMS implementation, ensuring the absence of issues like data races, deadlocks, and priority inversions. \cite{wang-etal-2022-iteratively} proposed an iterative prompting framework for pre-trained language models to handle multi-step reasoning tasks. It introduced a context-aware prompter that dynamically synthesised prompts based on the current step's context, improving the model's reasoning capabilities in complex tasks.

\section{Future Directions}
\label{sec:future_directions}

In this section, we outline prospective directions informed by the literature review. Much of the literature in this review employed queries containing a problem description and some instructions to achieve a desired outcome. Such querying of LLMs without training or examples of the current task is typically referred as zero-shot prompting and shows excellent performance on many tasks \cite{Kojima2022}.  Surprisingly, they also showed that the performance of LLM on some challenging problems can be improved by encouraging the LLM to reason using intermediate steps through a simple addition to problem prompts (“Lets think step by step”). 

\subsection{Advanced Prompt Engineering}
\label{subsec:CoTnPromptEngg}

Beyond this approach is one-shot prompting that includes an example of a solved problem to guide the LLM into generating the desired output \cite{li2024oneshotlearninginstructiondata}. This can be extended to few-shot prompting where a number of differing examples guide the LLM. But improved results are not assured as some studies e.g. \cite{NEURIPS2022c4025018} show that zero-shot can outperform the few-shot case \cite{zhang2023selfconvincedpromptingfewshotquestion}. \cite{Chen2024} reviewed the evolution of prompt engineering in LLMs, including discussions on self-consistency and multimodal prompt learning. It also reviewed the literature related to adversarial attacks and evaluation strategies for ensuring robust AI interactions.  

Chain of Thought (CoT) \cite{Jason2022} prompting involves a sequence of prompts producing intermediate results that are generated by the LLM and used to drive subsequent prompting interactions. These orchestrated interactions  can improve LLM performance on tasks requiring logic, calculation and decision-making in areas like math, common sense reasoning, and symbolic manipulation. CoT requires the LLM to articulate the distinct steps of its reasoning, by subdividing larger tasks into multi-step reasoning stages, acting as a precursor for subsequent stages.  But the CoT approach may require careful analysis when used with larger LLM offering long input contexts. This is because of the lost-in-the-middle problem where LLM show a U-shaped attention bias \cite{Chen2024} and can fail to attend to information in the middle of the context window. 

%\rmnote{ Add as part pof discussion: Emerging results investigating AI-driven methods for converting natural language into formal JML specifications using models like Mistral AI, OpenAI, Cohere, and Gemini. Here, two different strategies were tested: one using only the statement as input and another incorporating both the statement and corresponding code without specifications. While both approaches succeeded in some cases, challenges arose with loops and recursion, leading to the development of a dataset to address these difficulties.}

PromptCoT \cite{10656469} enhanced the quality of solutions for diffusion-based generative models by employing the CoT approach. The computational cost is minimised through adapter-based fine-tuning. Prompt design is explored in detail in \cite{amatriain2024promptdesignengineeringintroduction}. It discussed Chain-of-Thought and Reflection techniques, along with best practices for structuring prompts and building LLM-based agents.

Besta et al. \cite{Besta2024} introduced the concept of reasoning topologies, examining how structures such as Chains, Trees, and Graphs of Thought improve LLM reasoning. They also proposed a taxonomy of structured reasoning techniques, highlighting their influence on both performance and efficiency. Structured Chain-of-Thought (SCoT) prompting was proposed by \cite{10.1145/3690635} to enhance code generation by incorporating structured programming principles. This approach significantly improved the accuracy and robustness of LLM-based code synthesis compared to standard CoT methods. Building on the theme of automation, \cite{DBLP:conf/emnlp/ShumDZ23} introduced Automate-CoT, a technique for automatically generating and selecting rational chains for CoT prompting. By minimising dependence on human annotations, it enabled more flexible adaptation of CoT strategies across diverse reasoning tasks. Complementing these efforts, \cite{DBLP:journals/corr/abs-2302-11382} presented a prompt pattern catalog that offered reusable design patterns to optimise LLM interactions, thereby refining prompt engineering practices for a wide range of applications. Additionally, \cite{DBLP:journals/corr/abs-2305-09993} proposed Reprompting (Gibbs sampling-based algorithm) for discovering optimal CoT prompts. The proposed prompting technique consistently outperformed human-crafted alternatives and demonstrated high adaptability across various reasoning benchmarks.

Retrieval Augmented Generation (RAG) \cite{Lewis2020} supplements problem information with specifically retrieved information and is often used in knowledge intensive tasks. This helps ensure the LLM attends specifically to the retrieved information when addressing the users prompt. LLM model selection (chat vs reasoning) and fine tuning such as with LoRA \cite{Hu2022} remain among a growing number of possibilities for exploration.

Based on the literature survey conducted, we sketch one line research agenda: in VERIFAI, we aim to improve the techniques that bridge the gap between informal natural language description and rigorous formal specifications, through refinement of prompt engineering, the incorporation of chain-of-thought reasoning and the development of hybrid neuro-symbolic approaches.

\section{Conclusions}
\label{sec:conclusions}

The role of large language models in formalising software requirements is surveyed in this paper. The key contribution of the selected papers is on bridging the gap between informal natural language descriptions and rigorous formal specifications. We can enhance requirement formalisation through automating translations. The accuracy of translated ones is of key importance. We can deploy iterative refinements to improve the accuracy and correctness of the generated requirements. While LLMs is contributing significantly in improving development cycle, the challenges of ambiguity resolution, verification and domain adaptation shall be the focus areas of future research. 

In future research, the refinement of prompt engineering techniques, the incorporation of chain-of-thought reasoning and the development of hybrid neuro-symbolic approaches are the key areas to look at. Our survey concludes with the remarks that the better collaboration with industry and academia will further enhance the domain. 

\section{Acknowledgements}
  This work is partly funded by the ADAPT Research Centre for AI-Driven Digital Content Technology, which is funded by Research Ireland through the Research Ireland Centres Programme and is co funded under the European Regional Development Fund (ERDF) through Grant 13/RC/2106 P2.

\bibliographystyle{ACM-Reference-Format}
\bibliography{acmSurvBib}
\clearpage
\section{Appendix A - Summary Tables}

\begin{table}[ht]
    \centering
		\renewcommand{\arraystretch}{1.3}
    %\begin{adjustbox}{max width=\textheight} % Use textheight since it's landscape
    \begin{tabular}{|p{1cm}|p{3.5cm}|p{9cm}|}
    \hline
        \textbf{Paper} & \textbf{Tool / Framework / Technique} & \textbf{Description} \\
        \hline
        \cite{10500073} & LLM-based Code Verification & Uses LLMs like GPT-3.5 to verify code by analyzing requirements and explaining whether they are met. \\
        \hline
        \cite{cosler2023nl2specinteractivelytranslatingunstructured} & nl2spec & A framework leveraging LLMs to generate formal specifications from natural language, addressing ambiguity in system requirements with iterative refinement. \\
        \hline
        \cite{quan2024verificationrefinementnaturallanguage} & Explanation-Refiner & A neuro-symbolic framework integrating LLMs and theorem provers to formalize and validate explanatory sentences, providing error correction and feedback for improving NLI models. \\
        \hline
        \cite{10207159} & \textit{Not Specified} & Analyzes research directions in software requirement engineering, conducting a SWOT analysis and sharing evaluation findings. \\
        \hline
        \cite{10.1145/2976767.2976769} & Symbolic NLP vs. ChatGPT & Compares the performance of symbolic NLP and ChatGPT in generating correct JML output from natural language preconditions. \\
        \hline
        \cite{arora2023advancingrequirementsengineeringgenerative} & Domain Model Extractor & Generates domain models from natural language requirements in an industrial case study, evaluating accuracy and performance. \\
        \hline
        \cite{mandal2023largelanguagemodelsbased} & SpecSyn & A framework using LLMs for automatic synthesis of software specifications, improving accuracy by 21\% over previous tools. \\
        \hline
        \cite{10691792} & AssertLLM & A tool generating assertions for hardware verification from design specifications using three customized LLMs, achieving 89\% correctness. \\
        \hline
        \cite{10.1002/spe.430} & Formal Verification of NASA’s Software & Reports on formal verification of NASA's Node Control Software natural language specifications, highlighting errors and lessons learned. \\
        \hline
        \cite{li2024specllmexploringgenerationreview} & SpecLLM & Explores using LLMs for generating and reviewing VLSI design specifications, improving chip design documentation. \\
        \hline
        \cite{nowakowski2013requirements} & Requirements Specification Language (RSL), ReDSeeDS & Enhanced software requirements specification using constrained natural language and automated transformations into code. \\
    \hline
        \cite{Ghosh2016} & ARSENAL Framework and Methodology & Automated extraction of requirements specification from natural language with automatic verification. \\
   \hline
        \cite{6823180} & BPM-to-NL Translation Process & Generated natural language descriptions from business process models for better validation. \\
   \hline
        \cite{mugnier2024laurelgeneratingdafnyassertions} & Laurel & A framework to generate Dafny assertions to automate program verification process for a SMT solver \\ 
	 \hline
    \end{tabular}
    \caption{Summary of LLMs related literature: Tools, Frameworks, and Achievements}
    \label{tab:summary1to13}
\end{table}
\clearpage
\begin{table}[ht]
    \centering
		\renewcommand{\arraystretch}{1.3}
    %\begin{adjustbox}{max width=\textheight} % Use textheight since it's landscape
    \begin{tabular}{|p{1cm}|p{3.5cm}|p{9cm}|}
    \hline
        \textbf{Paper} & \textbf{Tool / Framework / Technique} & \textbf{Description} \\
        \hline
        \cite{mugnier2024laurelgeneratingdafnyassertions} & Laurel & A framework to generate Dafny assertions to automate program verification process for a SMT solver \\ 
	 \hline
        \cite{491451} & Controlled Natural Language (CL) with ANLT & Expressed software requirements in a limited set of natural language and translated to logical expressions to detect ambiguities. \\
   \hline
        \cite{reinpold2024exploringllmsverifyingtechnical} & LLM-based Analysis for Smart Grid Requirements & Improved smart grid requirement specifications with GPT-4o and Claude 3.5 Sonnet, achieving F1-scores between 79\% - 94\%. \\
   \hline
        \cite{10684640} & NL-to-LTL Translation via LLMs & Converted unstructured natural language requirements to NL-LTL pairs, achieving 94.4\% accuracy on public datasets. \\ \hline
        \cite{tihanyi2024newerasoftwaresecurity} & ESBMC-AI & Combined LLMs with Formal Verification to detect and fix software vulnerabilities with high accuracy. \\
   \hline
        \cite{hahn2022formalspecificationsnaturallanguage} & LLM-based Formal Specifications Translation & Translated natural language into formal rules (regex, FOL, LTL) with high adaptability and performance. \\
   \hline
        \cite{mukherjee2024automatedverificationllmsynthesizedc} & SynVer Framework & Synthesized and verified C programs using the Verified Software Toolchain. \\
   \hline
        \cite{10.1145/3643991.3644922} & LLM-based Requirement Coverage Analysis & Ensured low-level software requirements met high-level requirements, achieving 99.7\% recall in spotting missing coverage. \\
   \hline
    \cite{10.1145/3691620.3695302} & SAT-LLM & Integrated SMT solvers with LLMs to improve conflict identification in requirements; significantly outperformed standalone LLMs in detecting complex conflicts. \\ \hline
    \cite{ernst:LIPIcs.SNAPL.2017.4} & NLP for Software Development & Assessed NLP techniques for various software development stages, highlighting their suitability for generating assertions and processing developer queries. \\ \hline
    \cite{Req2SpecPaper} & Req2Spec & NLP-based tool that formalises natural language requirements for HANFOR; achieved 71\% accuracy in formalising 222 automotive requirements at BOSCH. \\ \hline
    \cite{Greenspan1986} & RML (Requirements Modeling Language) & Introduced a conceptual model-based framework ensuring precision, consistency, and clarity in requirements writing. \\ \hline
    \cite{fan2025evaluatingabilitylargelanguage} & GPT-4o for VeriFast Verification & Evaluated GPT-4o’s ability to generate C program specifications for VeriFast; found that while functional behavior was preserved, verification often failed or contained redundancies. \\ \hline
    \cite{Nelken1996} & NL to Temporal Logic Translation & Developed an automatic translation mechanism from natural language sentences to temporal logic for formal verification. \\ \hline
    \end{tabular}
    \caption{Summary of LLMs related literature: Tools, Frameworks, and Achievements}
    \label{tab:summary14to26}
\end{table}
\clearpage
\begin{table}[ht]
    \centering
		\renewcommand{\arraystretch}{1.3}
    %\begin{adjustbox}{max width=\textheight} % Use textheight since it's landscape
    \begin{tabular}{|p{1cm}|p{3.5cm}|p{9cm}|}
    \hline
        \textbf{Paper} & \textbf{Tool / Framework / Technique} & \textbf{Description} \\
        \hline
    \cite{Casadio2025} & ANTONIO toolkit & A comprehensive analysis of NLP verification approaches and introduces a structured \textbf{NLP Verification Pipeline} with six key components. The work includes identifying gaps in existing methods, proposing novel solutions for improved robustness, extending standard verifiability metrics, and emphasizing the importance of reporting verified subspace (geometric and semantic)  properties for better reliability and interpretability. \\ \hline
    \cite{wu2024lemurintegratinglargelanguage} & Lemur & Integrated LLMs with automated reasoners for program verification, defining sound transition rules and demonstrating improved performance on benchmark tests. \\ \hline
   % \cite{Si2020} & Code2Inv & Developed a reinforcement learning-based system to train an invariant synthesizer for end-to-end program verification. \\ \hline
    \cite{Necula2024} & Systematic Review & Conducted a comprehensive review on natural language to formal specification translation, analyzing research across multiple academic databases. \\ \hline
    \cite{10628478} & LLM-based Safety Requirements Pipeline & Designed a pipeline using LLMs to refine and decompose safety requirements for autonomous vehicles, evaluated through expert assessments and industrial implementation. \\ \hline
    \cite{7092662} & Specification Consistency Framework & Ensured consistency between oral and formal specifications, incorporating time extraction, input-output partitioning, and semantic reasoning, with positive evaluation results. \\ \hline
        \cite{10.5220/0006817205010512} & NLP Tools for TFM & Evaluated six NLP pipelines for Topological Functioning Modelling (TFM). Found that Stanford CoreNLP, FreeLing, and NLTK performed best. \\ \hline
        \cite{10.1145/3643763} & LLM-based Dafny Task Generation & Used LLMs (GPT-4, PaLM-2) to generate Dafny tasks from MBPP benchmark using different prompting strategies (context-less, signature, retrieval-augmented CoT). GPT-4 achieved best results with retrieval-augmented CoT prompt, producing 153 verified Dafny solutions. \\ \hline
        \cite{Yang2023} & LeanDojo \& ReProver & Introduced LeanDojo, an open-source toolkit for interacting with the Lean theorem prover. Developed ReProver, a retrieval-augmented LLM-based prover that improved theorem proving efficiency. Created a benchmark with 98,734 theorems and proofs for testing generalization. \\ \hline
       \cite{Jiang2022} & Thor and class methods named Hammers & Introduced a framework named Thor which integrates language models with theorem provers. Hammers are implemented to find the appropriate premises to complete the proofs of conjectures. Datasets used are PISA and MiniF2F. \\ \hline
       \cite{Granberry2025} & Not specified & The work proposes the integration of major formal languages (Dafny, Ada/SPARK, Frama-C, and KeY), their interactive theorem provers (Coq, Isabelle/HOL, Lean) with Copilot. \\ \hline 
       \cite{10.1145/3342355} & Systematic Review & Conducted a comprehensive survey on formal specification and verification of autonomous robotic systems in 2018. This is based on literature available of ten years (2008 - 2018). \\ \hline
      \cite{Granberry2025a} & Symbolic analysis and LLMs prompts & The quality of annotations produced in ACSL format is measured for PathCrawler and EVA (tools available in Frama-C). PathCrawler generated more context-aware annotations while, EVA efficiency improved having less run-time errors. \\ \hline
    \end{tabular}
    \caption{Summary of LLMs related literature: Tools, Frameworks, and Achievements}
    \label{tab:summary27to39}
\end{table}
\clearpage
\begin{table}[ht]
    \centering
		\renewcommand{\arraystretch}{1.3}
    %\begin{adjustbox}{max width=\textheight} % Use textheight since it's landscape
    \begin{tabular}{|p{1cm}|p{3.5cm}|p{9cm}|}
    \hline
    \textbf{Paper} & \textbf{Tool / Framework / Methodology Devised} & \textbf{Description} \\ \hline
    \cite{1618501} & Dynamic Requirements Traceability Model & Proposed a model to improve software quality through verification and validation of functional requirements, addressing scalability for both small and large projects. \\ \hline
    \cite{Salem2010} & Early Phase Traceability and Verification Model & Introduced a model for early development phase traceability and verification, with enhanced adaptability to requirement changes and impact analysis. \\ \hline
    \cite{Genvigir2010} & Comprehensive Review & Reviewed software requirements traceability, covering elements, challenges, techniques, classified approaches, and identified future research directions. \\ \hline
    \cite{7723818} & Empirical Analysis with Traceability Metrics & Conducted an empirical study enforcing requirements completeness, introducing traceability metrics and regression analysis to quantify software quality and reduce defect rates. \\ \hline
    \cite{Hayes2007} & RETRO & Developed RETRO, a tool for automating RTM generation, significantly improving accuracy and efficiency over manual tracing methods. \\ \hline
    \cite{doi:10.1142/S0218194020500278} & VSM + BTM-GA Hybrid Approach & Proposed a hybrid method for traceability link generation that outperformed traditional IR techniques, particularly in agile development, enhancing recall and precision. \\ \hline
    \cite{6341764} & Trustrace & Introduced a trust-based traceability recovery approach using mined repository data, achieving better precision and recall than standard IR methods. \\ \hline
    \cite{10.1109/ICSE.2017.9} & Deep Learning-Based Traceability (BI-GRU) & Applied deep learning with BI-GRU for traceability, incorporating semantic understanding and domain knowledge, outperforming VSM and LSI methods. \\ \hline
    \cite{DBLP:conf/enase/AsninaGOADS11} & Topology-Based Model-Driven Approach & Presented a topology-based, model-driven traceability technique formalising specifications and establishing trace links between real-world functions and software artifacts. \\ \hline
    \cite{1232285} & Event-Based Traceability Mechanism & Proposed a mechanism to support software evolution through event-based artifact linking, improving change management performance and maintaining consistency in distributed environments. \\ \hline
    \cite{7765529} & Tracebok & Introduced a traceability body of knowledge framework categorising traceability approaches and offering practical guidance for software projects. \\ \hline
    \cite{10.1007/978-981-16-8129-5_7} & Traceability Visualisation Review & Reviewed visualisation tools and techniques, identifying issues such as scalability and visual clutter, and suggesting ways to improve traceability visualisation. \\ \hline
    \end{tabular}
    \caption{Summary of Other Sections of Literature Review: Tools, Frameworks, and Methodologies}
    \label{tab:remaining_literature_summary1}
\end{table}
\clearpage
\begin{table}[ht]
    \centering
		\renewcommand{\arraystretch}{1.3}
    %\begin{adjustbox}{max width=\textheight} % Use textheight since it's landscape
    \begin{tabular}{|p{1cm}|p{3.5cm}|p{9cm}|}
    \hline
    \textbf{Paper} & \textbf{Tool / Framework / Methodology Devised} & \textbf{Description} \\ \hline
				    \cite{Kojima2022} & Zero-Shot CoT & Demonstrated that zero-shot prompts with simple additions like “Let’s think step by step” can significantly enhance LLM reasoning without training examples. \\ \hline
    \cite{li2024oneshotlearninginstructiondata} & One-Shot Prompting & Used a single example to guide LLMs in generating desired outputs, showing potential as an alternative to zero- and few-shot approaches. \\ \hline
    \cite{NEURIPS2022c4025018} & Few-Shot Evaluation & Found that zero-shot prompting can outperform few-shot setups, challenging assumptions on example-based prompting. \\ \hline
    \cite{zhang2023selfconvincedpromptingfewshotquestion} & Self-Consistent Few-Shot Prompting & Showed that prompting with few examples isn't always better and proposed techniques to refine few-shot reliability. \\ \hline
    \cite{Chen2024} & Prompt Engineering Review & Surveyed prompt engineering strategies, including multimodal prompts, adversarial prompting, and robustness evaluations. \\ \hline
    \cite{Jason2022} & Chain of Thought (CoT) & Proposed stepwise reasoning in prompts to improve LLM performance in tasks requiring logic, math, and symbolic manipulation. \\ \hline
    \cite{Chen2024} & Lost-in-the-Middle & Highlighted LLMs’ U-shaped attention patterns, warning against long prompts where important middle-context information may be overlooked. \\ \hline
    \cite{Lewis2020} & Retrieval Augmented Generation (RAG) & Introduced a method to retrieve relevant knowledge for augmenting prompts, enhancing performance on knowledge-intensive tasks. \\ \hline
    \cite{Hu2022} & LoRA & Proposed Low-Rank Adaptation for fine-tuning LLMs efficiently without retraining the entire model, supporting modular adaptability. \\ \hline
		\cite{wang-etal-2022-iteratively} & Iterative Prompting Framework & Proposed a context-aware iterative prompting method for LLMs in multi-step reasoning tasks, dynamically synthesizing prompts to improve reasoning accuracy. \\ \hline
    \cite{Woodcock2004} & UTP Tutorial & Provided an introduction to Unifying Theories of Programming (UTP) and the use of alphabetised relational calculus to describe imperative constructs such as Hoare logic and refinement calculus. \\ \hline
    \cite{10.1145/147508.147524} & Institutions & Introduced institutions as a formal framework for modeling logical systems, including results on signature gluing and constraints for abstract data types, contributing to the theory of specification languages. \\ \hline
    \cite{Woodcock2002} & Circus & Described Circus, a language combining CSP, Z, and imperative programming, formalised using UTP to refine concurrent systems. \\ \hline
    \cite{DBLP:journals/isse/HammerCHJR22} & Runtime Verification for UAS & Applied runtime verification in UAS traffic management, using formal requirements to validate system safety across subsystems, confirmed via flight simulations. \\ \hline
    \cite{DBLP:conf/birthday/ButterfieldT23} & UTP-Based RTEMS Verification & Verified RTEMS real-time OS using Promela and SPIN, linking UTP semantics for test generation and discussing future directions in space-grade verification. \\ \hline
    \end{tabular}
    \caption{Summary of Other Sections of Literature Review: Tools, Frameworks, and Methodologies}
    \label{tab:remaining_literature_summary2}
\end{table}
\clearpage
\begin{table}[ht]
    \centering
		\renewcommand{\arraystretch}{1.3}
    %\begin{adjustbox}{max width=\textheight} % Use textheight since it's landscape
    \begin{tabular}{|p{1cm}|p{3.5cm}|p{9cm}|}
    \hline
    \textbf{Paper} & \textbf{Tool / Framework / Methodology Devised} & \textbf{Description} \\ \hline
    \cite{FOSTER2020102510} & Isabelle/UTP & Introduced Isabelle/UTP to mechanise UTP semantics, providing formal proof tools for various paradigms, supporting development of automated verification tools. \\ \hline
    \cite{6058982} & UML Testing Profile (UTP) & Discussed use of UML and UTP for model-based testing of embedded systems, presenting algorithms for test artefact generation under resource constraints. \\ \hline
    \cite{Gadia2016} & Java Model of RTEMS & Presented a verified Java model for RTEMS's priority inheritance protocol using Java Pathfinder, fixing known issues like data races and deadlocks. \\ \hline
    \end{tabular}
    \caption{Summary of Other Sections of Literature Review: Tools, Frameworks, and Methodologies}
    \label{tab:remaining_literature_summary2}
\end{table}

\end{document}